%% file: Ereco_MachineLearning_Drakopoulou.tex
\documentclass[12pt]{article}
\usepackage{graphicx}
\usepackage{subfig}

\newcommand\pubnumber{NuPhys2016-Drakopoulou}
\newcommand\pubdate{\today}

\def\napoli{School of Physics and Astronomy\\
University of Edinburgh, EH9 3FD, Scotland, United Kingdom}

\def\Title#1{\begin{center} {\Large #1 } \end{center}}
\def\Author#1{\begin{center}{ \sc #1} \end{center}}
\def\Address#1{\begin{center}{ \it #1} \end{center}}

\newcommand\pubblock{\rightline{\begin{tabular}{l} \pubnumber\\
         \pubdate  \end{tabular}}}
\newenvironment{Abstract}{\begin{quotation}  }{\end{quotation}}
\newenvironment{Presented}{\begin{quotation} \begin{center} 
             PRESENTED AT \end{center}\bigskip 
      \begin{center}\begin{large}}{\end{large}\end{center} \end{quotation}}

\input econfmacros.tex

\begin{document}
\begin{titlepage}
\pubblock

\vfill
\Title{Machine Learning-based Energy Reconstruction for Water-Cherenkov detectors}
\vfill
\Author{ Greig Cowan, Evangelia Drakopoulou, Matthew Needham, Mahdi Taani}
\Address{\napoli}
\vfill
\begin{Abstract}
Hyper-Kamiokande (Hyper-K) is a proposed next generation underground water Cherenkov (WCh) experiment. The far detector will measure the oscillated neutrino flux from the long-baseline neutrino experiment using 0.6 GeV neutrinos produced by a 1.3 MW proton beam at J-PARC. It has a broad program of physics and astrophysics mainly focusing on the precise measurement of the lepton neutrino mixing matrix and the CP asymmetry. The unoscillated neutrino flux will be measured by an intermediate WCh detector. One of the proposed designs is the Tokai Intermediate Tank for the Unoscillated Spectrum (TITUS). WCh detectors are instrumented with photomultipliers to detect the Cherenkov light emitted from charged particles which are produced by neutrino interactions. The detection of light is used to measure the energy, position and direction of the charged particles. We propose machine learning-based methods to reconstruct the energy of charged particles in WCh detectors and present our results for the TITUS configuration.
\end{Abstract}
\vfill
\begin{Presented}
NuPhys2016, Prospects in Neutrino Physics\\
Barbican Centre, London, UK,  December 12-14, 2016
\end{Presented}
\vfill
\end{titlepage}

\section{Introduction}
The Hyper-K experiment is a proposed 0.5 Mton water Cherenkov detector which will act as the far detector for the future long-baseline neutrino program in Japan \cite{hyperk}. It will measure the oscillated spectrum of the neutrino beam produced by the proton synchrotron in J-PARC. The detector will be placed at 2.5$^{\circ}$ off the beam axis, resulting in a narrow neutrino beam profile peaked at the P($\nu_{\mu} \rightarrow \nu_{e}$) oscillation maximum of 0.6 GeV at 295 km. The main focus of the experiment is the precise measurement of the leptonic CP asymmetry. To minimise uncertainties due to the modelling of neutrino interactions and hence the neutrino flux, an intermediate detector, TITUS \cite{titus}, using the same target material as the far detector has been proposed. 
This is a 2 kton water Cherenkov detector aiming to measure the neutrino spectrum before oscillation. Water Cherenkov detectors detect neutrino interactions using the photons emitted when the charged particles traverse the water volume. The Cherenkov photons are collected by photomultipliers (PMTs) and are used to reconstruct the energy, position and direction of the charged particles. Previously, the energy reconstruction of the charged particles for the TITUS detector was performing using look up tables \cite{titus}. We propose a new approach using machine learning-based methods for the energy reconstruction.

\section{Machine Learning Energy Reconstruction}

Machine learning methods are gaining popularity as quick and efficient tools in multidisciplinary problems. A wide variety of algorithms ranging from neutral networks to boosted decision trees are implemented in several software packages. For this study, we compared neural networks with boosted decision trees using an appropriate selection of input variables for the energy reconstruction of charged particles in water Cherenkov detectors. Here, we present the two best options of machine learning algorithms for our problem. These are boosted decision trees with Gradient Boost (BDTG) implemented in the ROOT-TMVA \cite{tmva} and the Scikit \cite{scikit} package. This method was applied to both muons and electrons produced by $\nu_{\mu}$ and $\nu_{e}$ events respectively with comparable performance. 


\subsection{Input Variables}

The boosted decision trees are trained using appropriate input variables and the Monte Carlo (MC) muon energy. The weights produced for each variable during the training phase are used for the estimation of the muon energy. The input variables which are selected for the energy reconstruction are: \\
i) The number of hits in PMTs: The number of photoelectrons (pes) in PMTs are clustered according to time coincidences under the Cherenkov hypothesis and the total number of selected hits in clusters is measured.\\
ii) The number of photoelectrons in rings: To take into account the possibility that in high energies more than one particles produce Cherenkov rings, the observed photoelectrons are grouped in rings. Then, the total number of photoelectrons for the observed rings is measured. \\ 
iii) The vertical distances from the reconstructed track direction to the detector walls \\
iv) The track length calculated as the distance between the first and last PMT hit under the hypothesis of Cherenkov emission angle.

\begin{figure}[htb]
\centering
 \centering
  \includegraphics[height=2.in]{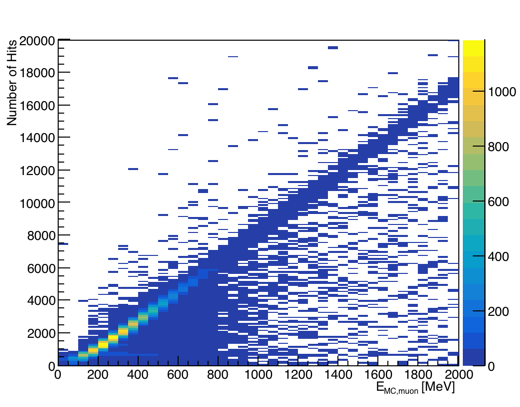}
 \centering
  \includegraphics[height=2.in]{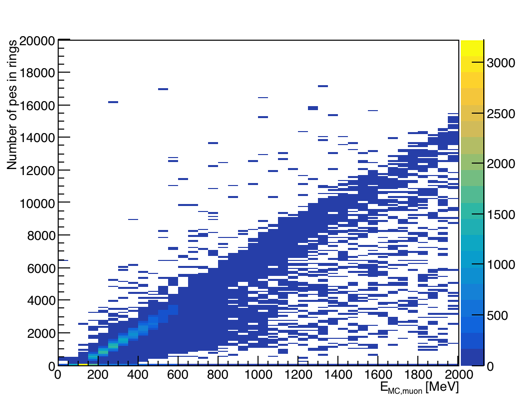}
\caption{The number of hits in PMTs (left plot) and the number of photoelectrons in rings (right plot) as a function of the MC muon energy.}
\label{fig:hits_emu}
\end{figure}

Variables (i) and (ii) have a strong dependence on the MC muon energy as it is shown in shown in Figure ~\ref{fig:hits_emu}. The last pair of variables is used to take into account the event topology, thus compensating for events that escape the instrumented volume.

\subsection{Results on the Energy Reconstruction}

\begin{figure}[htb]
\centering
 \centering
  \includegraphics[height=2.in]{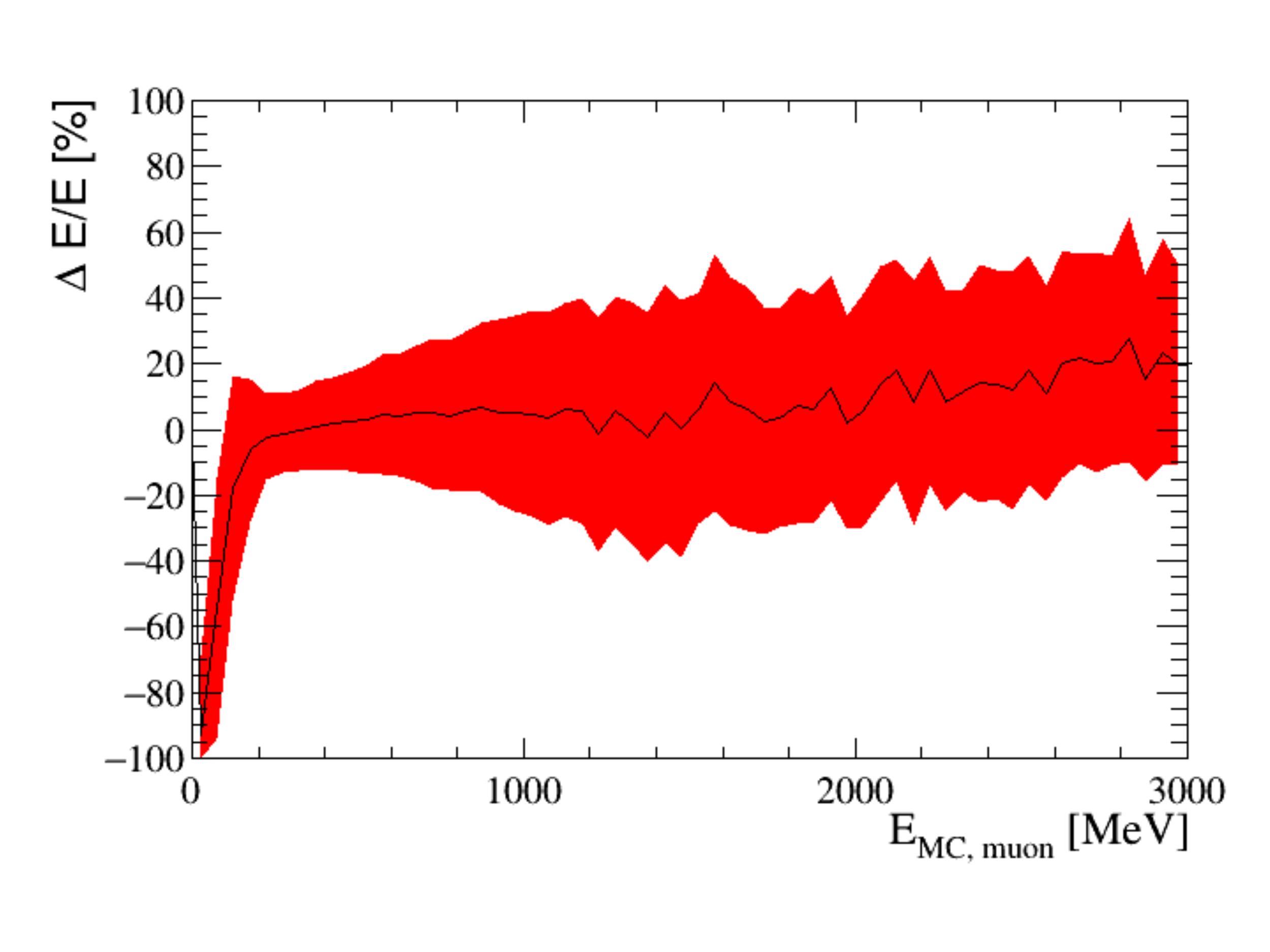}
 \centering
  \includegraphics[height=2.in]{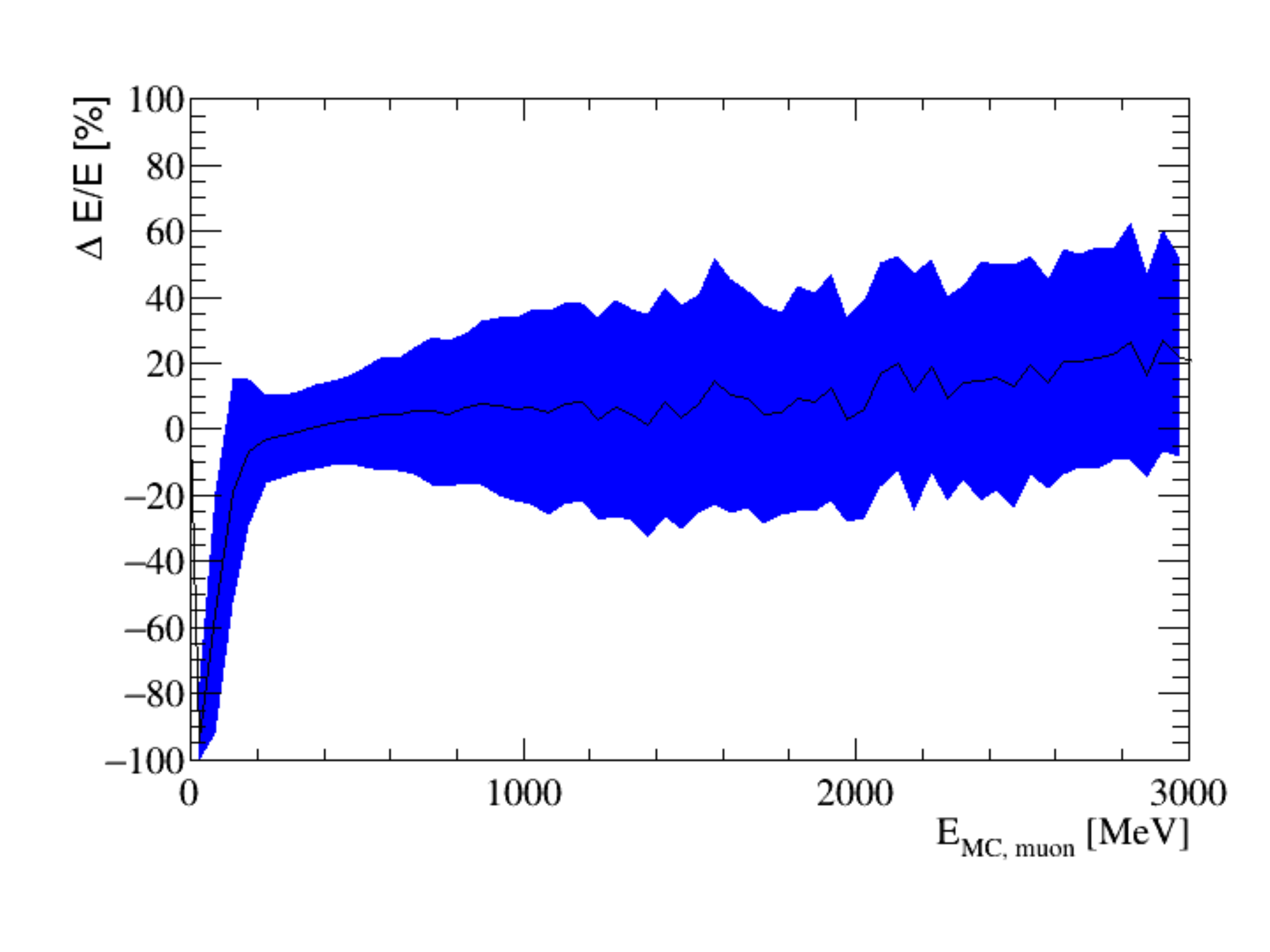}
\caption{The mean and standard deviation of the resolution distribution in bins of energy for the BDTG from the TMVA (left plot) and the Scikit (right plot) package.}
\label{fig:res1}
\end{figure}

The BDTG algorithms from the TMVA and the Scikit package were trained and tested for muon events using the input variables described in Section 2.1. The results of the energy resolution as a function of the MC muon energy are shown in Figure ~\ref{fig:res1}. The energy resolution is defined as: 
\begin{center}
$\frac{{\textrm{$\Delta$E}}}{\textrm{E}}=\frac{\textrm{MC muon energy - Reconstructed muon energy}}{\textrm{MC muon energy}}\times{100}$
\end{center} 

\begin{figure}[htb]
\centering
  \includegraphics[height=2.in]{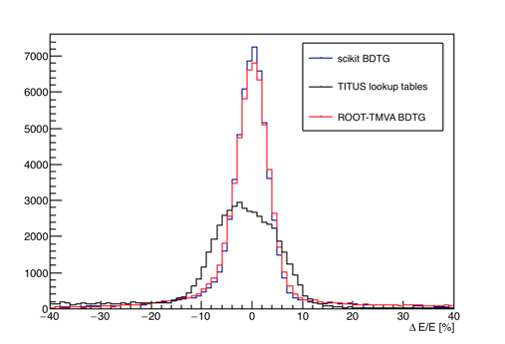}
  \caption{The  distribution of the energy resolution for muon energies from 200 MeV to 600 MeV for  the TMVA and the Scikit package and for the TITUS look up tables.}
  \label{fig:res200600}
\end{figure}

Both methods have comparable performance resulting in a very good energy reconstruction for muon energies from 200 MeV to 600 MeV which is the area of interest. This is shown in Figure ~\ref{fig:res200600}. Both methods lead to a  better energy resolution compared to the TITUS method using look up tables with analogous input variables. The use of the BDTG from the Scikit package leads to a lower standard deviation and therefore better energy reconstruction for high energy muons (with energies above 2 GeV). Both methods will be further optimised and tested with other approaches in the future.

\begin{figure}[htb]
\centering
 \centering
  \includegraphics[height=2.in]{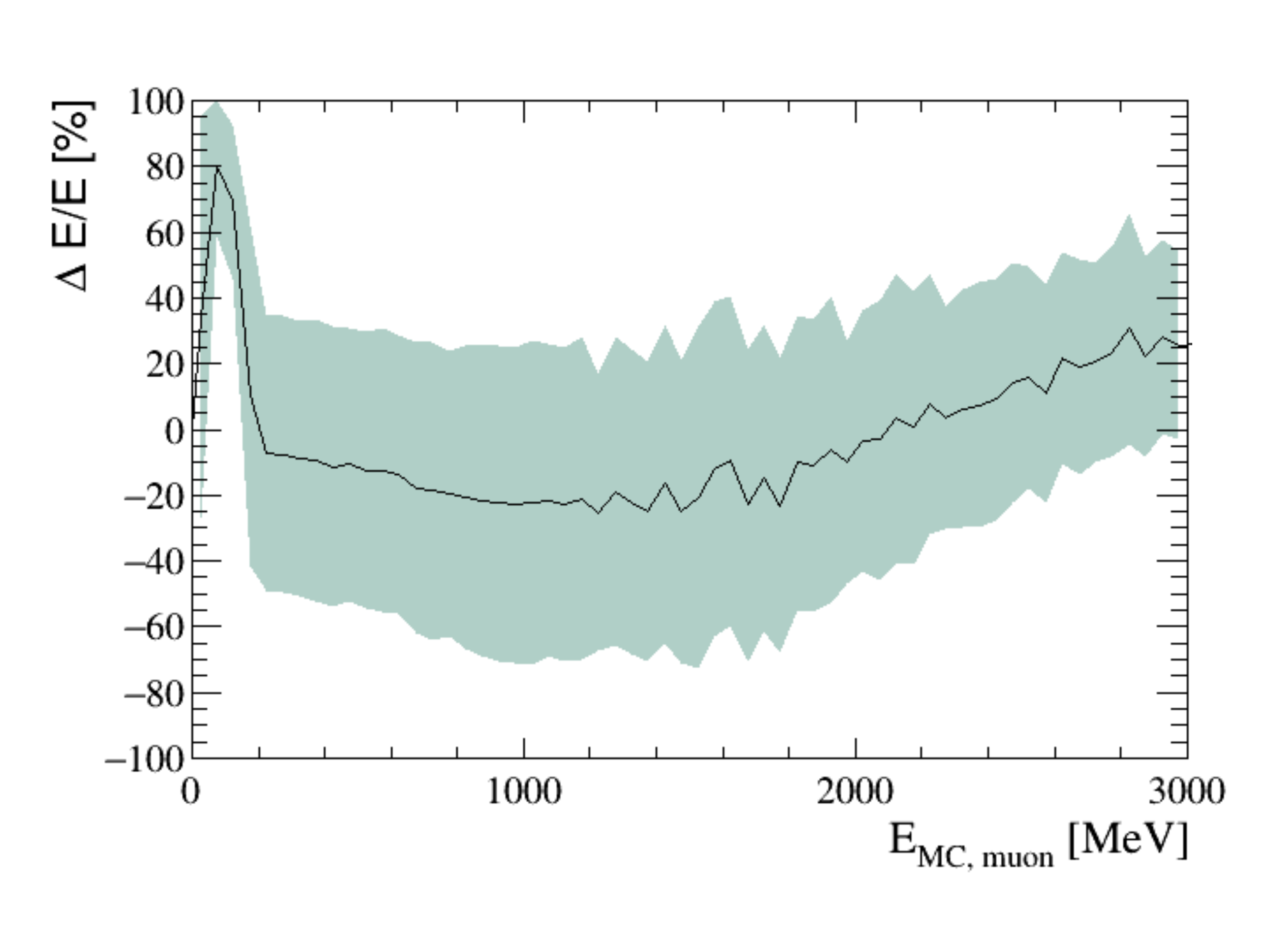}
 \centering
  \includegraphics[height=2.in]{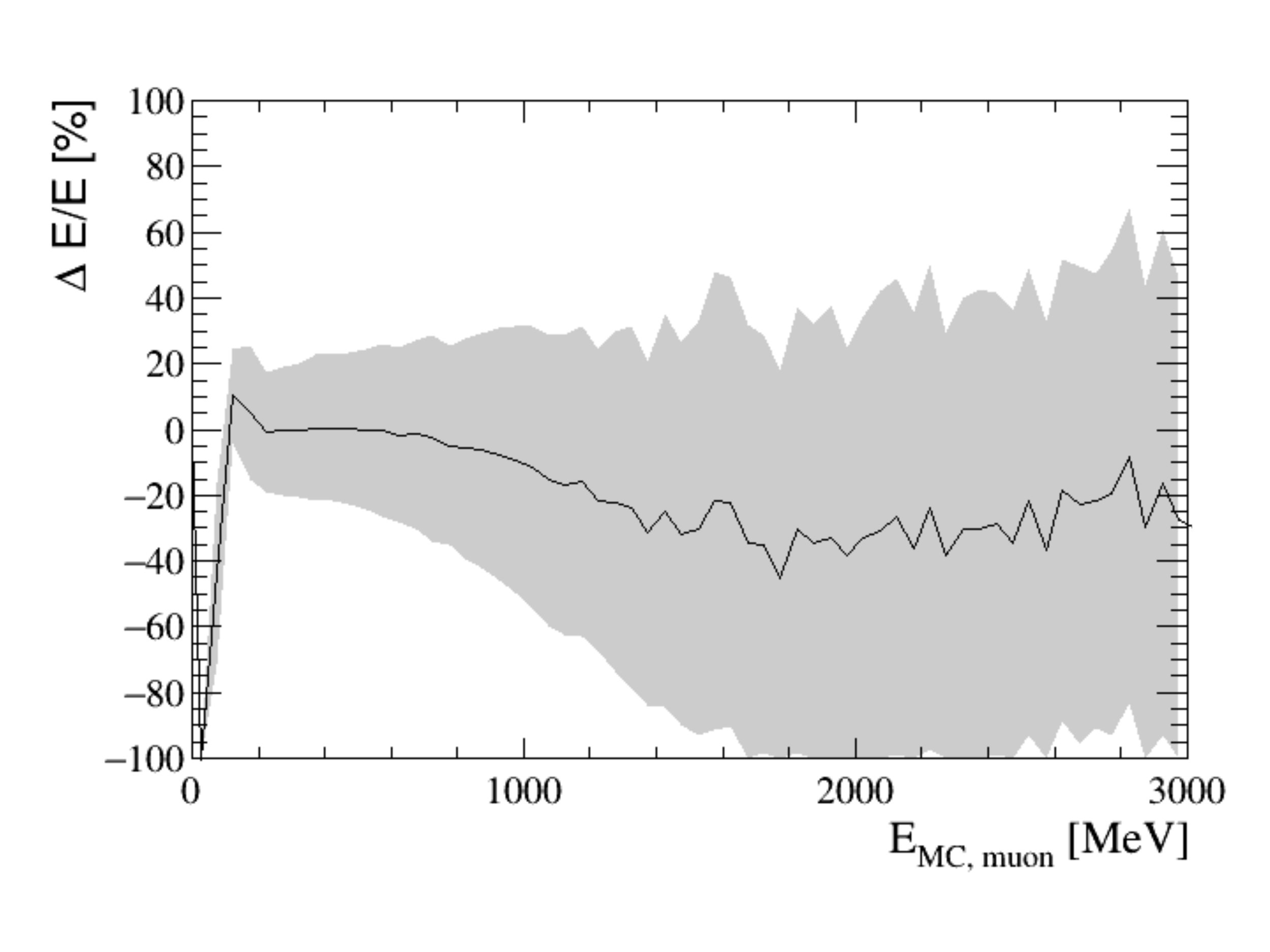}
\caption{The mean and standard deviation of the resolution distribution in bins of energy for the TITUS look up tables (left plot) and the new look up tables (right plot).}
\label{fig:res2}
\end{figure}

In Figure ~\ref{fig:res2} (left) the muon energy resolution using the TITUS look up tables with analogous input variables is shown. This approach leads to larger deviations in the energy resolution across the whole muon energy range. 
\par To  understand why the results on the energy reconstruction with TITUS look up tables are worse than the machine learning approach we created new look up tables using only the number of photoelectrons in rings. As it is shown in Figure ~\ref{fig:res2} (right), these look up tables have better performance for muon energies below 1 GeV. This is due to the fact that the TITUS look up tables assume a spherical event topology instead of a cylindrical one. When this bias is eliminated and only the number of photoelectrons is used their performance is improved. However, both look up tables result in worse energy resolution compared to the BDTGs.

\begin{figure}[htb]
\centering
 \centering
  \includegraphics[height=2.in]{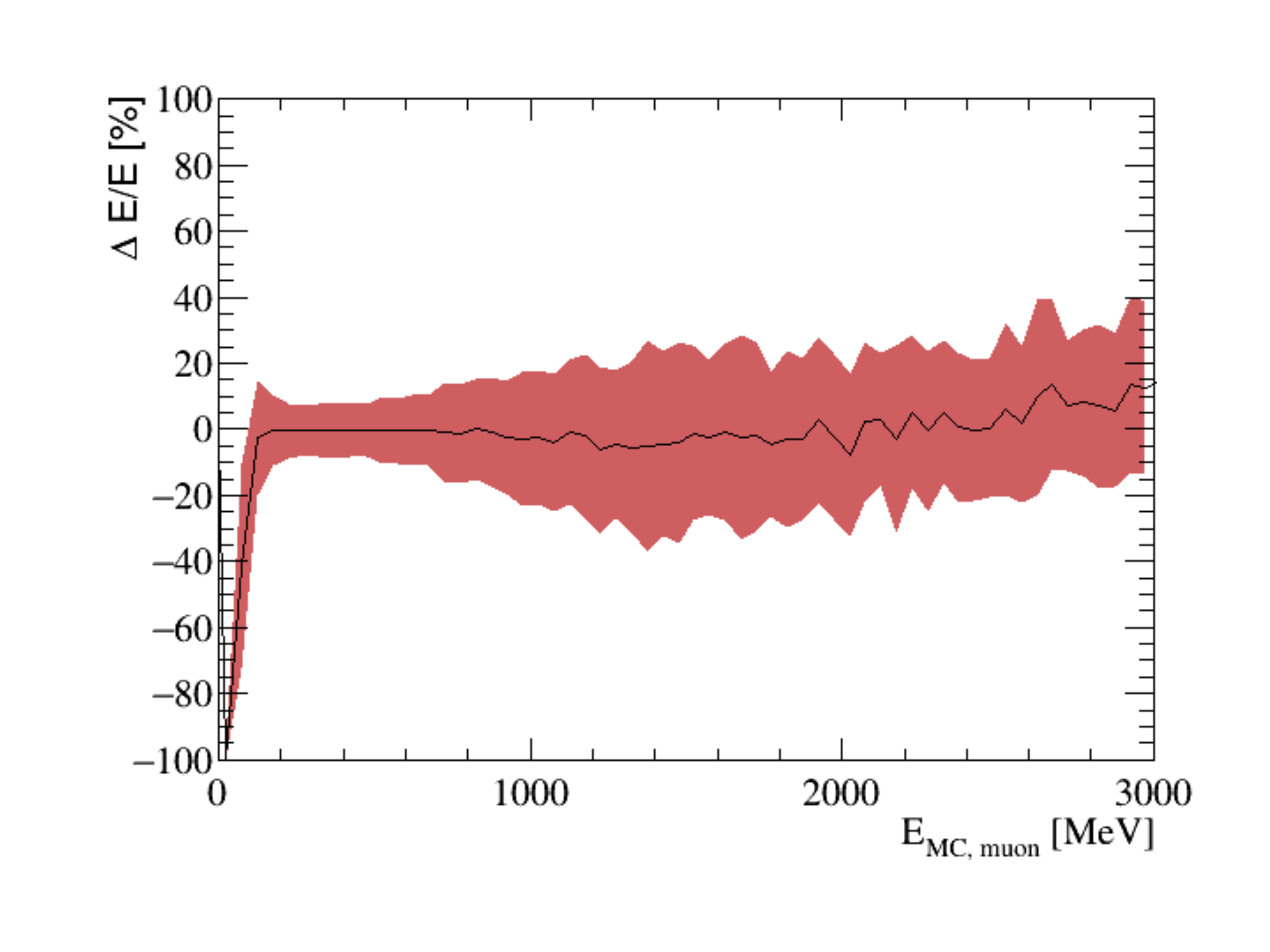}
 \centering
  \includegraphics[height=2.in]{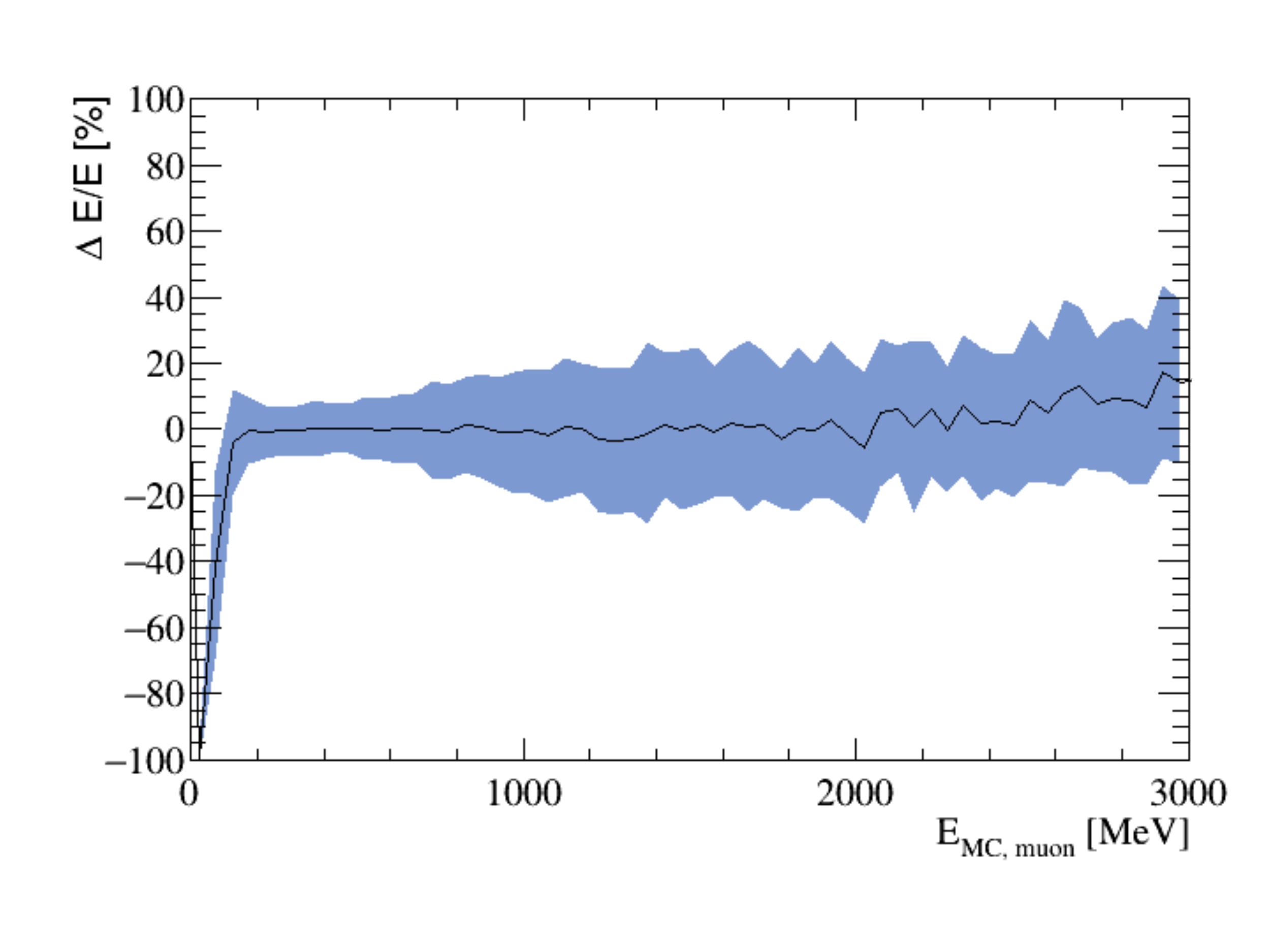}
\caption{The mean and standard deviation of the resolution distribution in bins of energy for the BDTG from the TMVA (left plot) and the Scikit (right plot) package for events in a fiducial volume.}
\label{fig:resINfid}
\end{figure}

For the results mentioned above, all simulated events have been used. However, for detailed analysis studies, only events in a fiducial volume are taken into account. These are events that are well-contained in the detector and deposit most of their energy in the instrumented volume. The energy resolution for events in a cylindrical fiducial volume 1 m away from each detector wall is shown in Figure ~\ref{fig:resINfid} for the TMVA (left) and the Scikit (right) BDTG.  Both methods lead to a very good energy resolution for this subsample of events across the whole muon energy range.

\section{Conclusion}

A new method for the energy reconstruction using BDTGs with appropriate input variables from ROOT-TMVA and Scikit packages was presented. The results for both BDTGs are comparable and better than the standard TITUS approach using look up tables, thus improving the standard deviation of the energy resolution. 
This method can be used to reconstruct both the muon and electron energy, thus constituting a new technique for the energy reconstruction in water Cherenkov detectors. 


\end{document}

%% file: econfmacros.tex
\def\beq{\begin{equation}}
\def\eeq#1{\label{#1}\end{equation}}
\def\eeqn{\end{equation}}

\def\beqa{\begin{eqnarray}}
\def\eeqa#1{\label{#1}\end{eqnarray}}
\def\eeqan{\end{eqnarray}}

\let\bar=\overbar

\def\Dslash{\not{\hbox{\kern-4pt $D$}}}
\def\dslash{\not{\hbox{\kern-2pt $\del$}}}

\def\msb{{\bar{\ssstyle M \kern -1pt S}}}

